\documentclass[%
 reprint,
superscriptaddress,
 amsmath,amssymb,
 aps,
 prl,
]{revtex4-2}

\usepackage{graphicx}
\usepackage{siunitx}
\usepackage{ulem}
\usepackage{xcolor}
\usepackage{supertabular}
\usepackage{booktabs} 
\usepackage{threeparttable}
\usepackage{makecell}
\usepackage{subcaption}
\usepackage{lineno}
\usepackage[colorlinks=true]{hyperref}

\makeatletter
\renewcommand\@makecaption[2]{%
  \par
  \vskip\abovecaptionskip
  \begingroup
   \small\rmfamily
    \begingroup
     \samepage
     \flushing
     \let\footnote\@footnotemark@gobble
     \@make@capt@title{#1}{#2}\par
    \endgroup
  \endgroup
  \vskip\belowcaptionskip
}
\makeatother
\begin{document}

\title{Measurement of Solar $pp$ Neutrino Flux using Electron Recoil Data from PandaX-4T Commissioning Run}

\def\shKeyLab{School of Physics and Astronomy, Shanghai Jiao Tong University, Key Laboratory for Particle Astrophysics and Cosmology (MoE), Shanghai Key Laboratory for Particle Physics and Cosmology, Shanghai 200240, China}
\def\scKeyLab{Jinping Deep Underground Frontier Science and Dark Matter Key Laboratory of Sichuan Province}
\def\BUAA{School of Physics, Beihang University, Beijing 102206, China}
\def\BUAALab{Beijing Key Laboratory of Advanced Nuclear Materials and Physics, Beihang University, Beijing 102206, China}
\def\BUAAPHC{Peng Huanwu Collaborative Center for Research and Education, Beihang University, Beijing 100191, China}
\def\SCNT{Southern Center for Nuclear-Science Theory (SCNT), Institute of Modern Physics, Chinese Academy of Sciences, Huizhou 516000, China}
\def\USTClab{State Key Laboratory of Particle Detection and Electronics, University of Science and Technology of China, Hefei 230026, China}
\def\USTCdep{Department of Modern Physics, University of Science and Technology of China, Hefei 230026, China}
\def\pku{School of Physics, Peking University, Beijing 100871, China}
\def\YaLongSD{Yalong River Hydropower Development Company, Ltd., 288 Shuanglin Road, Chengdu 610051, China}
\def\IAP{Shanghai Institute of Applied Physics, Chinese Academy of Sciences, 201800 Shanghai, China}
\def\CHEPpku{Center for High Energy Physics, Peking University, Beijing 100871, China}
\def\SDUdep{Research Center for Particle Science and Technology, Institute of Frontier and Interdisciplinary Science, Shandong University, Qingdao 266237, Shandong, China}
\def\SDUlab{Key Laboratory of Particle Physics and Particle Irradiation of Ministry of Education, Shandong University, Qingdao 266237, Shandong, China}
\def\UMD{Department of Physics, University of Maryland, College Park, Maryland 20742, USA}
\def\TDLee{New Cornerstone Science Laboratory, Tsung-Dao Lee Institute, Shanghai Jiao Tong University, Shanghai, 200240, China}

\def\MESJTU{School of Mechanical Engineering, Shanghai Jiao Tong University, Shanghai 200240, China}
\def\SYU{School of Physics, Sun Yat-Sen University, Guangzhou 510275, China}
\def\SYUSFI{Sino-French Institute of Nuclear Engineering and Technology, Sun Yat-Sen University, Zhuhai, 519082, China}
\def\NKU{School of Physics, Nankai University, Tianjin 300071, China}
\def\YTU{Department of Physics,Yantai University, Yantai 264005, China}
\def\FDU{Key Laboratory of Nuclear Physics and Ion-beam Application (MOE), Institute of Modern Physics, Fudan University, Shanghai 200433, China}
\def\USST{School of Medical Instrument and Food Engineering, University of Shanghai for Science and Technology, Shanghai 200093, China}
\def\SJTUSC{Shanghai Jiao Tong University Sichuan Research Institute, Chengdu 610213, China}
\def\SPEIT{SJTU Paris Elite Institute of Technology, Shanghai Jiao Tong University, Shanghai, 200240, China}
\def\NNU{School of Physics and Technology, Nanjing Normal University, Nanjing 210023, China}
\def\SYUzhuhai{School of Physics and Astronomy, Sun Yat-Sen University, Zhuhai, 519082, China}

\author{Xiaoying Lu}\affiliation{\SDUdep}\affiliation{\SDUlab}
\author{Abdusalam Abdukerim}\affiliation{\shKeyLab}
\author{Zihao Bo}\affiliation{\shKeyLab}
\author{Wei Chen}\affiliation{\shKeyLab}
\author{Xun Chen}\affiliation{\TDLee}\affiliation{\shKeyLab}\affiliation{\SJTUSC}\affiliation{\scKeyLab}
\author{Yunhua Chen}\affiliation{\YaLongSD}\affiliation{\scKeyLab}
\author{Chen Cheng}\affiliation{\SYU}
\author{Zhaokan Cheng}\affiliation{\SYUSFI}
\author{Xiangyi Cui}\affiliation{\TDLee}
\author{Yingjie Fan}\affiliation{\YTU}
\author{Deqing Fang}\affiliation{\FDU}
\author{Lisheng Geng}\affiliation{\BUAA}\affiliation{\BUAALab}\affiliation{\BUAAPHC}\affiliation{\SCNT}
\author{Karl Giboni}\affiliation{\shKeyLab}\affiliation{\scKeyLab}
\author{Xuyuan Guo}\affiliation{\YaLongSD}\affiliation{\scKeyLab}
\author{Chencheng Han}\affiliation{\TDLee} 
\author{Ke Han}\email[Corresponding author: ]{ke.han@sjtu.edu.cn}\affiliation{\shKeyLab}\affiliation{\scKeyLab}
\author{Changda He}\affiliation{\shKeyLab}
\author{Jinrong He}\affiliation{\YaLongSD}
\author{Di Huang}\affiliation{\shKeyLab}
\author{Junting Huang}\affiliation{\shKeyLab}\affiliation{\scKeyLab}
\author{Zhou Huang}\affiliation{\shKeyLab}
\author{Ruquan Hou}\affiliation{\SJTUSC}\affiliation{\scKeyLab}
\author{Yu Hou}\affiliation{\MESJTU}
\author{Xiangdong Ji}\affiliation{\UMD}
\author{Yonglin Ju}\affiliation{\MESJTU}\affiliation{\scKeyLab}
\author{Chenxiang Li}\affiliation{\shKeyLab}
\author{Jiafu Li}\affiliation{\SYU}
\author{Mingchuan Li}\affiliation{\YaLongSD}\affiliation{\scKeyLab}
\author{Shuaijie Li}\affiliation{\YaLongSD}\affiliation{\shKeyLab}\affiliation{\scKeyLab}
\author{Tao Li}\affiliation{\SYUSFI}
\author{Qing Lin}\affiliation{\USTClab}\affiliation{\USTCdep}
\author{Jianglai Liu}\email[Spokesperson: ]{jianglai.liu@sjtu.edu.cn}\affiliation{\TDLee}\affiliation{\shKeyLab}\affiliation{\SJTUSC}\affiliation{\scKeyLab}
\author{Congcong Lu}\affiliation{\MESJTU}
\author{Lingyin Luo}\affiliation{\pku}
\author{Yunyang Luo}\affiliation{\USTCdep}
\author{Wenbo Ma}\affiliation{\shKeyLab}
\author{Yugang Ma}\affiliation{\FDU}
\author{Yajun Mao}\affiliation{\pku}
\author{Yue Meng}\affiliation{\shKeyLab}\affiliation{\SJTUSC}\affiliation{\scKeyLab}
\author{Xuyang Ning}\affiliation{\shKeyLab}
\author{Binyu Pang}\affiliation{\SDUdep}\affiliation{\SDUlab}
\author{Ningchun Qi}\affiliation{\YaLongSD}\affiliation{\scKeyLab}
\author{Zhicheng Qian}\affiliation{\shKeyLab}
\author{Xiangxiang Ren}\affiliation{\SDUdep}\affiliation{\SDUlab}
\author{Nasir Shaheed}\affiliation{\SDUdep}\affiliation{\SDUlab}
\author{Xiaofeng Shang}\affiliation{\shKeyLab}
\author{Xiyuan Shao}\affiliation{\NKU}
\author{Guofang Shen}\affiliation{\BUAA}
\author{Manbin Shen}\affiliation{\YaLongSD}\affiliation{\scKeyLab}
\author{Lin Si}\affiliation{\shKeyLab}
\author{Wenliang Sun}\affiliation{\YaLongSD}\affiliation{\scKeyLab}
\author{Yi Tao}\affiliation{\shKeyLab}\affiliation{\SJTUSC}
\author{Anqing Wang}\affiliation{\SDUdep}\affiliation{\SDUlab}
\author{Meng Wang}\affiliation{\SDUdep}\affiliation{\SDUlab}
\author{Qiuhong Wang}\affiliation{\FDU}
\author{Shaobo Wang}\affiliation{\shKeyLab}\affiliation{\SPEIT}\affiliation{\scKeyLab}
\author{Siguang Wang}\affiliation{\pku}
\author{Wei Wang}\affiliation{\SYUSFI}\affiliation{\SYU}
\author{Xiuli Wang}\affiliation{\MESJTU}
\author{Xu Wang}\affiliation{\TDLee}
\author{Zhou Wang}\affiliation{\TDLee}\affiliation{\shKeyLab}\affiliation{\SJTUSC}\affiliation{\scKeyLab}
\author{Yuehuan Wei}\affiliation{\SYUSFI}
\author{Mengmeng Wu}\affiliation{\SYU}
\author{Weihao Wu}\affiliation{\shKeyLab}\affiliation{\scKeyLab}
\author{Yuan Wu}\affiliation{\shKeyLab}
\author{Mengjiao Xiao}\affiliation{\shKeyLab}
\author{Xiang Xiao}\affiliation{\SYU}
\author{Kaizhi Xiong}\affiliation{\YaLongSD}\affiliation{\scKeyLab}
\author{Binbin Yan}\email[Corresponding author: ]{yanbinbin@sjtu.edu.cn}\affiliation{\TDLee}
\author{Xiyu Yan}\affiliation{\SYUzhuhai}
\author{Yong Yang}\affiliation{\shKeyLab}\affiliation{\scKeyLab}
\author{Chunxu Yu}\affiliation{\NKU}
\author{Ying Yuan}\affiliation{\shKeyLab}
\author{Zhe Yuan}\affiliation{\FDU} %
\author{Youhui Yun}\affiliation{\shKeyLab}
\author{Xinning Zeng}\affiliation{\shKeyLab}
\author{Minzhen Zhang}\affiliation{\TDLee}
\author{Peng Zhang}\affiliation{\YaLongSD}\affiliation{\scKeyLab}
\author{Shibo Zhang}\affiliation{\TDLee}
\author{Shu Zhang}\affiliation{\SYU}
\author{Tao Zhang}\affiliation{\TDLee}\affiliation{\shKeyLab}\affiliation{\scKeyLab}
\author{Wei Zhang}\affiliation{\TDLee}
\author{Yang Zhang}\affiliation{\SDUdep}\affiliation{\SDUlab}
\author{Yingxin Zhang}\affiliation{\SDUdep}\affiliation{\SDUlab} %
\author{Yuanyuan Zhang}\affiliation{\TDLee}
\author{Li Zhao}\affiliation{\TDLee}\affiliation{\shKeyLab}\affiliation{\scKeyLab}
\author{Jifang Zhou}\affiliation{\YaLongSD}\affiliation{\scKeyLab}
\author{Ning Zhou}\affiliation{\TDLee}\affiliation{\shKeyLab}\affiliation{\SJTUSC}\affiliation{\scKeyLab}
\author{Xiaopeng Zhou}\affiliation{\BUAA}
\author{Yubo Zhou}\affiliation{\shKeyLab}
\author{Zhizhen Zhou}\affiliation{\shKeyLab}
\collaboration{PandaX Collaboration}
\noaffiliation

\date{\today}
\begin{abstract}
The proton-proton ($pp$) fusion chain dominates the neutrino production from the Sun. 
The uncertainty of the predicted $pp$ neutrino flux is at the sub-percent level, whereas that of the best measurement is $\mathcal{O}(10\%)$. 
In this paper, we present the first result to measure the solar $pp$ neutrinos in the electron recoil energy range from 24 to 144 keV, using the PandaX-4T commissioning data with 0.63 tonne$\times$year exposure.
The $pp$ neutrino flux is determined to be $(8.0 \pm 3.9 \,{\rm{(stat)}} \pm 10.0 \,{\rm{(syst)}} )\times 10^{10}\, $$\rm{s}^{-1} \rm{cm}^{-2}$, consistent with Standard Solar Model and existing measurements, corresponding to a flux upper limit of $23.3\times 10^{10}\, $$\rm{s}^{-1} \rm{cm}^{-2}$ at 90\% C.L.. 
\end{abstract}

\maketitle

According to the Standard Solar Model (SSM)~\cite{Adelberger:1998qm,Turck-Chieze:2001aug,Adelberger:2010qa}, approximately 99\% of solar power comes from a series of reactions fusing hydrogen into helium.
Neutrinos emitted in the initial fusion of two protons to a deuteron constitute roughly 91\% of the total solar neutrino flux. These neutrinos are commonly referred to as the $pp$ neutrinos, with a theoretical flux uncertainty at the sub-percent level. 
The sub-dominant neutrinos from electron capture of $^7$Be account for an additional 7\% of the flux.
The precise measurement of solar neutrino flux is essential for verifying the SSM and shedding light on the solar metallicity puzzle~\cite{Vinyoles:2016djt}.
Solar neutrinos are also crucial for neutrino physics, especially for understanding the matter effect of neutrino oscillation~\cite{deHolanda:2002dko}.
The first observation of $pp$ neutrino flux is achieved using radiochemical material of gallium by GALLEX/GNO~\cite{GALLEX:1992gcp} and SAGE~\cite{SAGE:1994ctc}. The first real-time detection is made by Borexino with now a state-of-the-art precision of $\mathcal{O}(10\%)$ and an electron recoil energy threshold of 165~keV~\cite{BOREXINO:2014pcl, BOREXINO:2018ohr}.
In this paper, we present a new result to measure $pp$ neutrino flux in real-time by PandaX-4T within a previously unaccessible recoil energy window between 24 to 144 keV. It also represents the first such measurement using a detector primarily designed for dark matter direct detection. 

A number of multi-tonne-scale liquid xenon (LXe) experiments, including PandaX-4T~\cite{PandaX-4T:2021bab}, LZ~\cite{LZ:2018qzl}, and XENONnT~\cite{XENON:2020kmp} are under operation to search for dark matter particles, coherent scattering of solar neutrinos on xenon nuclei~\cite{PandaX:2022aac,XENON:2020gfr}, and possible abnormal magnetic moments of neutrinos~\cite{PandaX-II:2020udv,XENON:2020rca,Billard:2013qya} in the few or few-tens of keV-scale electron or nuclear recoil energy.
By design, these detectors also effectively cover an electron recoil energy up to several hundred keV, spanning over most of the energy region of the $pp$ neutrino.

\begin{figure}[tbp]
    \centering

        \includegraphics[width=\columnwidth]{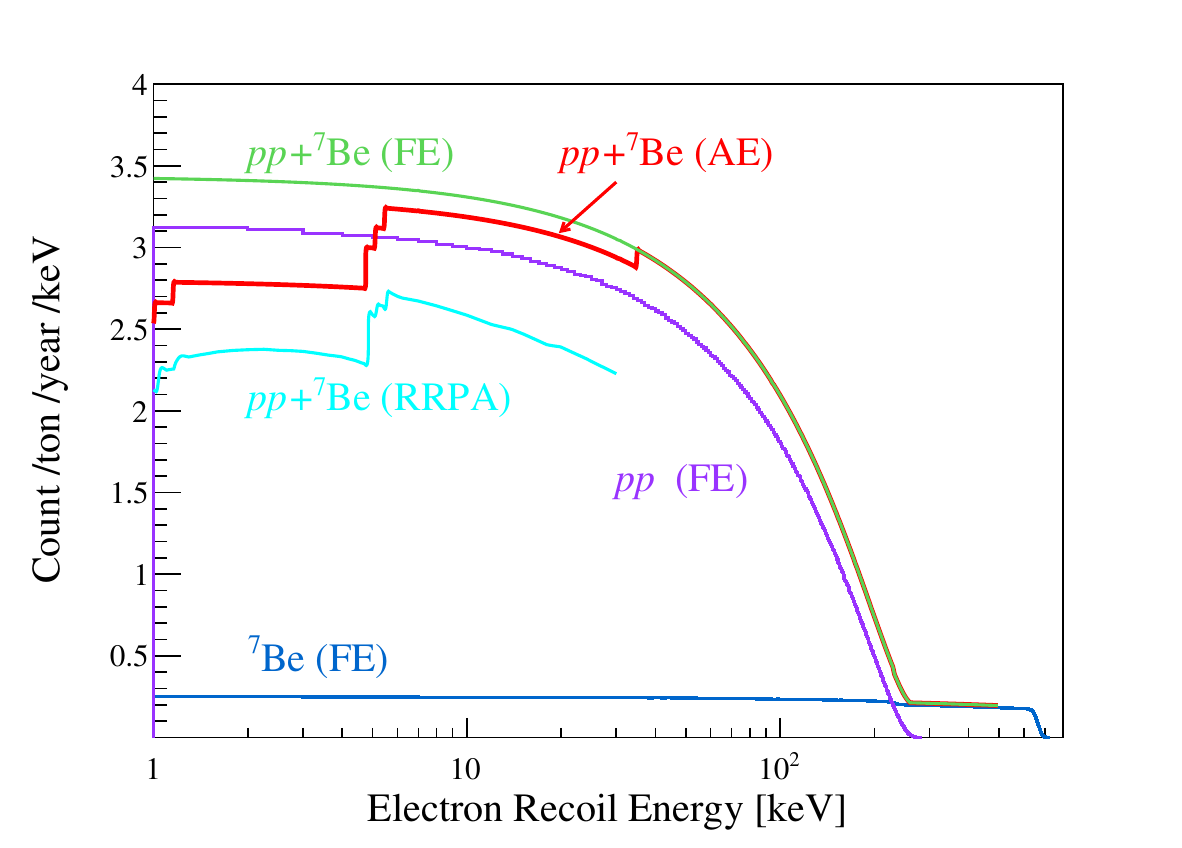}
        \caption{Neutrino-electron elastic scattering spectrum of solar neutrinos with liquid xenon. 
        If the binding energy of atomic electrons (AE) is considered~\cite{Chen:2016eab} (red curve), the scattering event rate in our region of interest is lower than the free electron (FE) scenario (green curve). The more complete treatment using the relativistic random phase approximation (RRPA) is shown in cyan, but only for a recoil energy less than 30~keV. 
        }
        \label{fig:fig1_ppLXe}
\end{figure}

The solar $pp$ neutrinos have a continuous energy spectrum with an endpoint at 420~keV. 
The $^{7}$Be neutrinos are mono-energetic at 384~keV (approximately 10\%) and 862~keV (90\%).
The neutrino and electron interact via the electroweak force through the exchange of a Z or W boson, the latter of which is only possible for an electron neutrino.
The expected electron recoil event rate per unit recoil energy is
\begin{equation}
    \begin{aligned}
        \frac{d R}{d E_r} &=N \sum_j \int \phi(E_\nu) P_{e j} \frac{d \sigma_j(E_\nu,E_r)}{d E_r} d E_\nu \,,
    \end{aligned}
\end{equation}
where $N$ is the number of target electrons, $\phi(E_\nu)$ is the neutrino flux as a function of neutrino energy, $P_{ej}$ ($j=e,\mu,\tau$) is the oscillation probabilities of $\nu_e$ into flavor j and $d \sigma_j$ is the differential cross-section. 
Fig.~\ref{fig:fig1_ppLXe} shows the electron recoil energy spectrum of solar $pp$ and $^7$Be neutrinos in a LXe detector, as given in Ref.~\cite{Chen:2016eab}.
When the xenon atomic effects are considered (adopted in this paper), the rate in the region of interest (ROI) is suppressed by a few percent compared to the free electron scenario. 

The PandaX-4T detector is located in the B2 hall of the China Jinping Underground Laboratory~\cite{CDEX:2017kys,Li:2014rca}.
The sensitive target of PandaX-4T contains 3.69~tonne of liquid xenon in a cylindrical dual-phase xenon time projection chamber (TPC) with 118.5~cm in diameter and 116.8~cm in height~\cite{PandaX:2023ggs}. 
The prompt scintillation photons ($S1$) and the delayed electroluminescence photons ($S2$) are produced when a given energy is deposited in the sensitive volume.
Both $S1$ and $S2$ signals are recorded by the top and bottom photomultipliers (PMT) arrays, which have 169 and 199 Hamamatsu 3-inch PMTs installed.
Detailed discussions of the PandaX-4T detector can be found in Ref.~\cite{PandaX-4T:2021bab}.

\begin{figure}[tbp]
    \centering
    \includegraphics[width=0.5\textwidth]{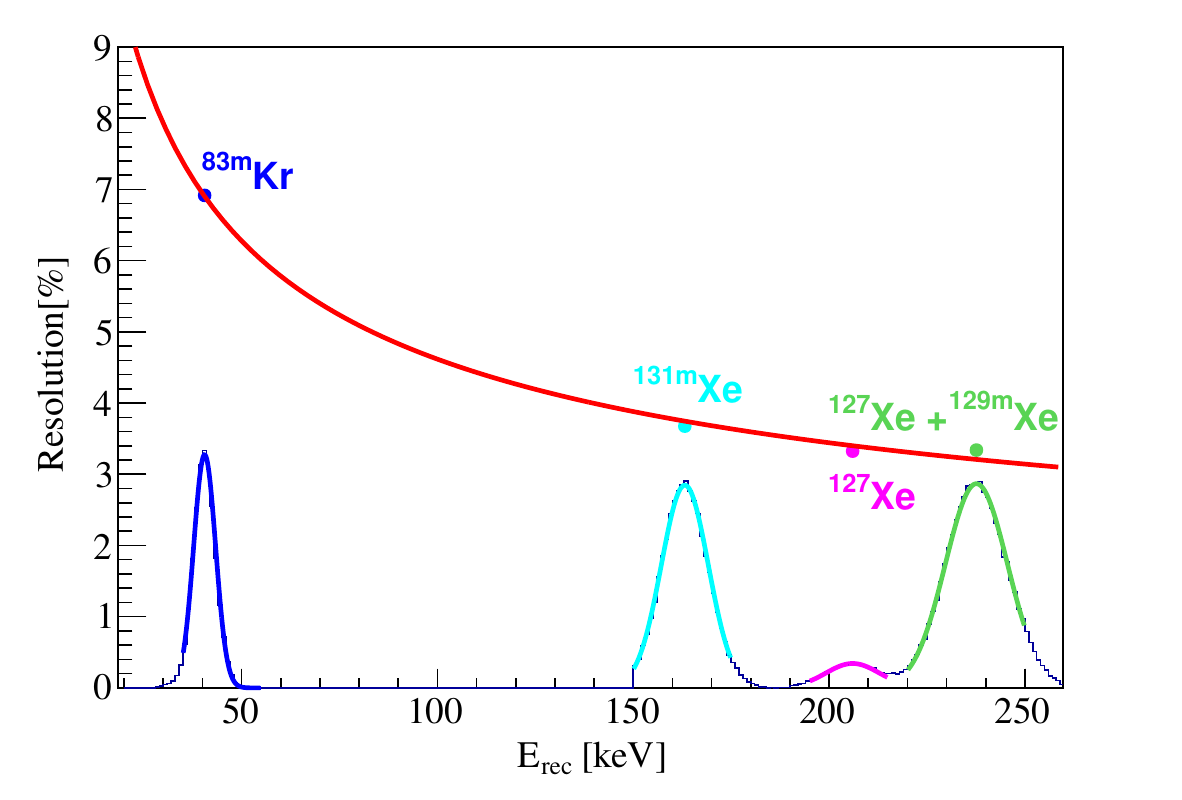}
    \caption{The energy resolution using characteristic peaks at 41.6~keV, 163.9~keV, 208.1~keV, and 236.1~keV. 
    The red curve shows the fit function $\sigma/E = 0.40/\sqrt{E\,[\mathrm{keV}]}$ + 0.0061.}
    \label{fig:energySpectrum}
\end{figure}

This analysis selects different data samples compared to existing analyses of the commissioning data release~\cite{PandaX-4T:2021bab,PandaX:2022aac,PandaX:2022kwg}.
The energy ROI is from 24~keV to 144~keV, with more than 60\% of the solar $pp$ neutrino events included.
The lower boundary is chosen to be above the dark matter particle search region and the upper boundary is to avoid the prominent 163.9~keV peak from $^{\rm{131m}}$Xe produced in the neutron calibration runs.
In the ROI, the detector noise has a negligible effect on both $S1$ and $S2$ signals.
Therefore, we can recover approximately 9.5\% of exposure previously excluded in the dark matter analysis due to elevated detector noise~\cite{PandaX-4T:2021bab}.
On the contrary, 8.4~days of data following a neutron calibration run is removed because of a high concentration of activated $^{133}$Xe and $^{125}$I.
In the end, a total data set of 86.5 days is used for this analysis.

The PMT gains are calibrated with a newly implemented ``rolling gain'' approach, which fits single photoelectron (SPE) spectra in individual PMTs run by run, adopted already in recent Ref.~\cite{PandaX:2023ggs}.
Our previous analyses calculated PMT gains with weekly light-emitting-diode (LED) calibration.
To avoid biases in our data selections, selection cuts are first determined with LED-gain calibration data, validated with about 13.5 days of rolling gain physics data, and finally applied to the entire data set.

The quality cut variables are inherited from the dark matter analysis~\cite{PandaX-4T:2021bab}, but cut parameters are modified 
to suit the differing energy window.
The main difference is the relaxation of the top and bottom PMT charge ratio requirement in the $S2$ signal selection to accommodate the top $S2$ saturation. 
The total quality cut efficiency is $(99.1 \pm 0.1)\%$.
The scattering of solar $pp$ neutrinos on electrons is primarily single-site (SS).
The identification of SS events follows the same procedure as in Ref.~\cite{PandaX-4T:2021bab}.
The SS efficiency is $(99.7 \pm 0.1)\%$, calculated using the $^{220}$Rn calibration and consistent with simulation. 

The horizontal position reconstruction follows the same procedure in Ref.~\cite{PandaX:2023ggs}, where de-saturated waveforms and improved optical Monte Carlo simulation are used.
The reconstruction uniformity is confirmed with the diffusive $^{83\rm m}$Kr calibration source injected into the TPC~\cite{Zhang:2021shp}.
The mono-energetic peak of $^{83\rm m}$Kr is also used to generate an energy response map. The corresponding correction procedure follows Ref.~\cite{PandaX-4T:2021bab}.

The energy in LXe TPC is reconstructed as in Refs.~\cite{Szydagis:2011tk,XENON:2020iwh}.
The energy spectrum is further corrected using a quadratic function between reconstructed energy and true energy of characteristic peaks at 41.6~keV ($\mathrm{^{83m}Kr}$), 163.9~keV ($\mathrm{^{131m}Xe}$), 208.1~keV ($\mathrm{^{127}Xe}$), and 236.1~keV ($\mathrm{^{129m}Xe}$ and $\mathrm{^{127}Xe}$).
After the correction, the residual offset of energy peaks in the ROI is smaller than 1~keV, which is considered as the systematic uncertainty of the energy reconstruction.
The relative $1\sigma$ energy resolution at 24~keV (144~keV) is 8.8\% (3.9\%), as shown in Fig.~\ref{fig:energySpectrum}.

\begin{figure}[tbp]
    \centering
        \includegraphics[width=0.9\columnwidth]{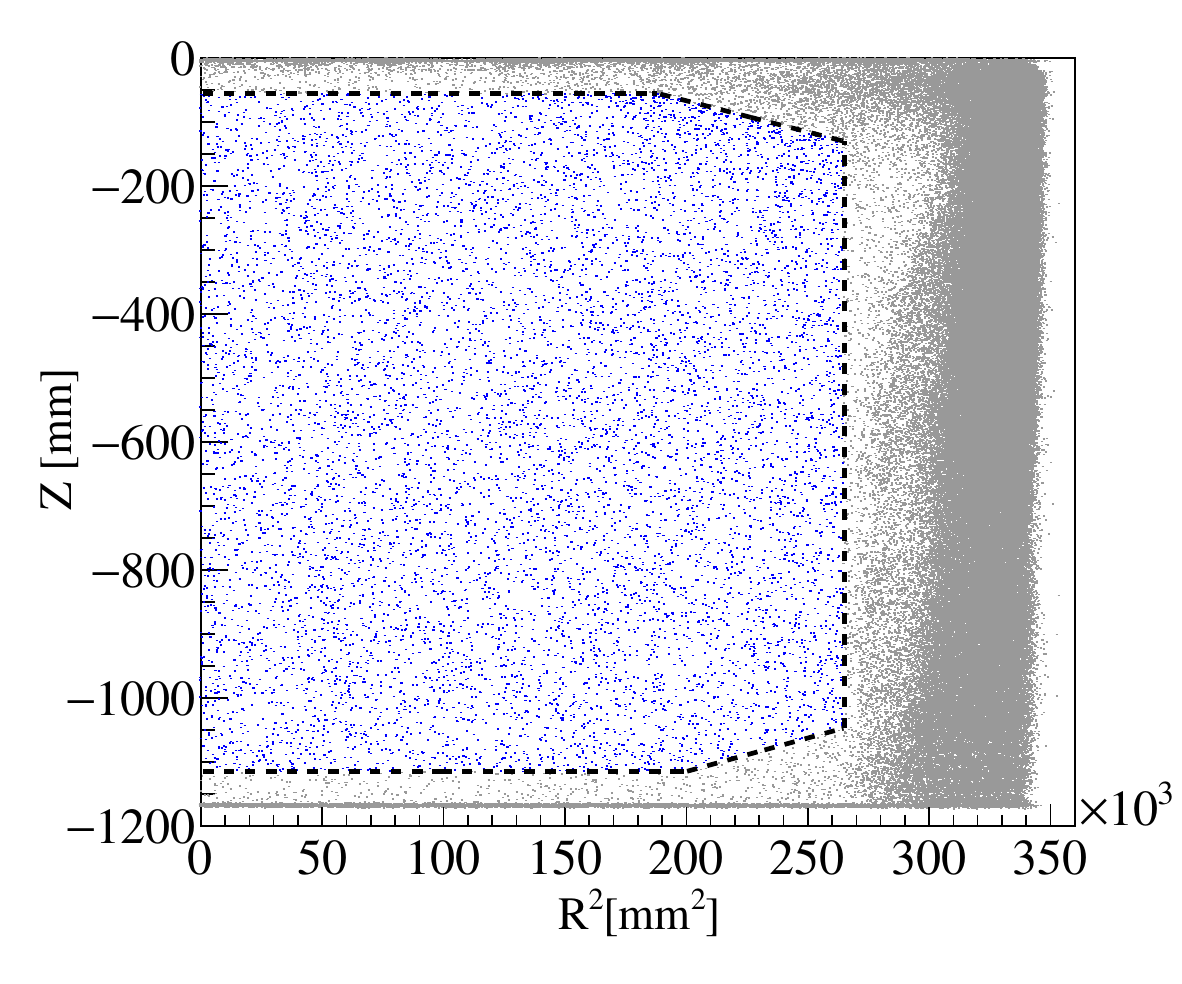}
        \includegraphics[width=0.9\columnwidth]{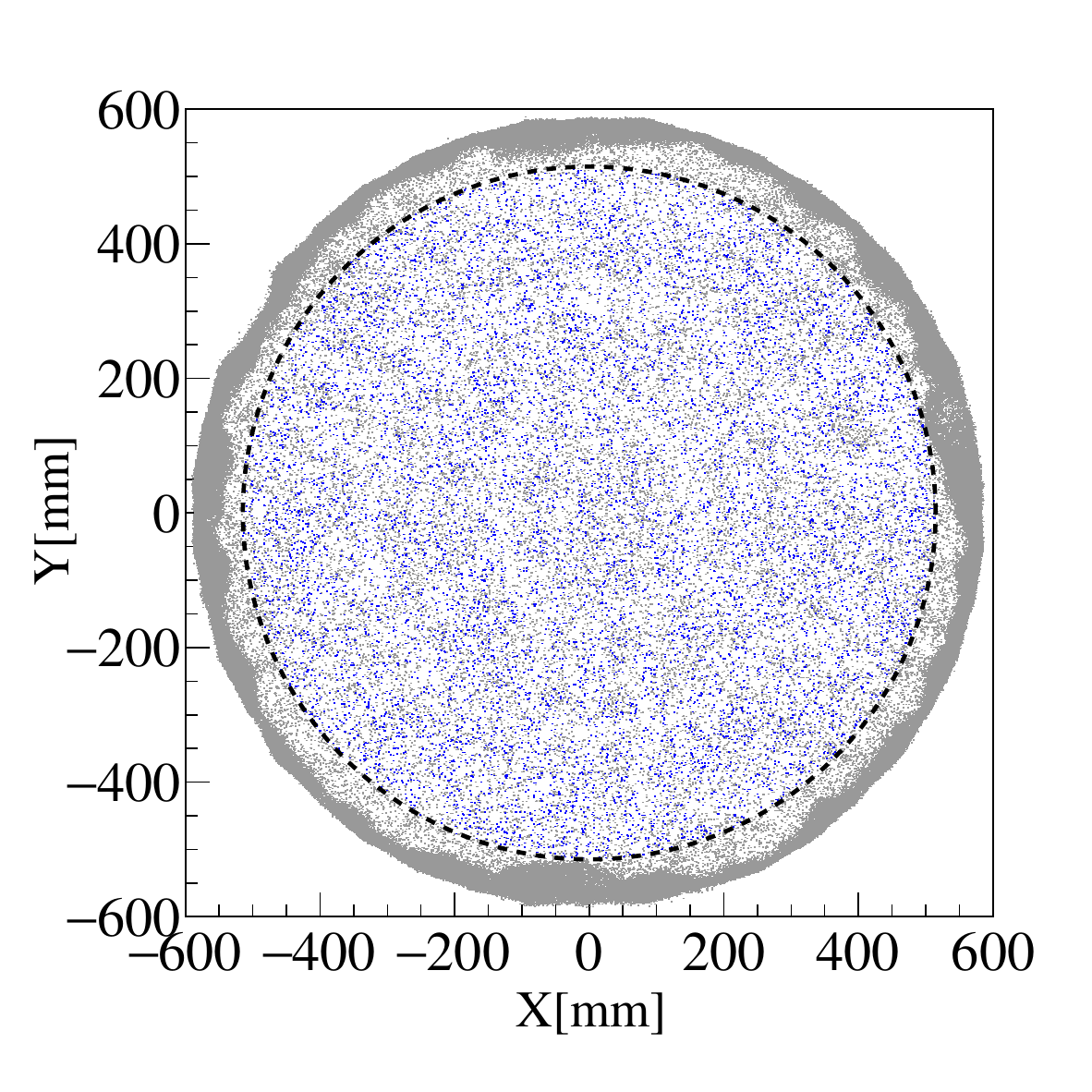}
    \caption{The spatial distributions of the selected physics events in $ \mathrm Z$ vs. $\mathrm R^2$ (top) and $\mathrm Y$ vs. $\mathrm X$ (bottom).
    The dashed lines show the boundary of the fiducial volume, and blue (gray) dots represent events inside (outside) the fiducial volume.}
    \label{fig:fig_pos}
\end{figure}

The fiducial volume (FV) cuts and events in the ROI are shown in Fig.~\ref{fig:fig_pos} as dashed lines and blue dots, respectively. 
The FV selection is the same as in the dark matter analysis geometrically, and a total of $2.66\pm0.02$~tonne of LXe in the center of the TPC is used.
The slight difference with respect to Ref.~\cite{PandaX-4T:2021bab} is from the event position reconstruction.

The primary challenge in this analysis lies in the accurate rate estimation of various background components with smooth energy spectra that resemble the shape of the solar $pp$ signals.
The background sources in the ROI include radon and krypton impurities in liquid xenon, detector materials, and radioactive xenon isotopes.
The expected background sources are listed in Table~\ref{tab:solar_nu_background} and described below.

One major background in the ROI arises from the progeny of the internal $^{222}$Rn -- the $\beta$ decay of $^{214}$Pb to the ground state of $^{214}$Bi~\cite{PandaX:2023ggs}.
The ratio of the events in the ROI to the full SS energy spectrum of $^{214}$Pb $\beta$ decay is determined by a dedicated $^{222}$Rn injection calibration run, with $3.5\times 10^5$ $^{214}$Pb SS events accumulated over approximately 11~days.
The event ratio is measured to be $0.039 \pm 0.001$, with the uncertainty estimated based on the difference between our data and a recent theoretical prediction to accommodate the non-unique first-forbidden nature of the $\mathrm{^{214}Pb}$ decay~\cite{Haselschwardt:2020iey}.
The corresponding number of $^{214}$Pb events in the ROI is $1865\pm110$, in which the uncertainty combines both the fitted uncertainty of the overall $^{214}$Pb decay rate in the physics data~\cite{PandaX:2023ggs} and the uncertainty of the event ratio within our ROI. Also along the decay chain of $^{222}$Rn, the $\beta$ decay of $^{214}$Bi can be identified effectively using the $^{214}$Bi--$^{214}$Po coincidence, as the half-life of $^{214}$Po is 164~$\mu$s.
In our analysis, when a $^{214}$Bi-like event is found, we reject the event if an $\alpha$ event is found within the following 5~ms window. The residual $^{214}$Bi is no longer included in the spectrum fit. The loss of live time due to random coincidence is negligible. 

The expected contribution from the $\beta$ decay of $^{212}$Pb in the $^{220}$Rn decay chain follows the approach in Ref.~\cite{PandaX:2023ggs}. 
The ratio between $^{212}$Pb and the subsequent $^{212}$Po $\alpha$ events is determined using a $^{220}$Rn injection calibration run. Then using the $^{212}$Po $\alpha$ events identified in the physics data, 
$^{212}$Pb events in the ROI is estimated to be $276 \pm 71$, where the uncertainty is dominated by the variation of $^{212}$Pb/$^{212}$Po ratio during the $^{220}$Rn injection run.

The concentration of $^{85}$Kr is estimated to be 0.52 $\pm$ 0.27 parts per trillion~\cite{PandaX:2023ggs}, leading to 489 $\pm$ 254 events in the ROI.
The levels of radioactivity from PMTs and detector vessels have been determined from the wide energy spectrum fit in Ref.~\cite{PandaX:2022kwg}, resulting in $683 \pm 27$ events. The expected background from $^{136}$Xe $2\nu\beta\beta$ is $1009 \pm 46$ in the ROI, given by the half-life measured by the PandaX-4T experiment with an exposure of 15.5 kg$\times$year of $^{136}$Xe isotope~\cite{PandaX:2022kwg}. 

The contributions of $^{133}$Xe, $^{127}$Xe, $^{125}$I, and $^{124}$Xe are free and fitted in the final spectral analysis, as they cannot be constrained by our data outside the ROI.
The energy spectrum of $^{133}$Xe, activated during a neutron calibration run, has a distinctive rising slope starting at 81~keV. 
Cosmogenic $^{127}$Xe, introduced when some xenon exposed at the Earth's surface was filled into the detector, contributes to the background with K-shell electron capture around 33.2~keV.
Neutron-activated $^{125}$Xe decays quickly to 
the relatively long-lived $^{125}$I with a half-life of 59.4~days. 
$^{125}$I electron capture can release a total energy of 67.3~keV (K-shell), 40.4~keV (L-shell), and 36.5~keV (M-shell)~\cite{XENON:2022evz}. The reduction rate of $^{125}$I observed in the TPC is significantly more rapid than the natural lifetime of $^{125}$I, likely due to the removal by the xenon circulation and purification system.
Two-neutrino double-electron capture of $^{124}$Xe (abundance $\eta$= 0.095\%, Q = 2857~keV) deposits energy within our ROI at 64.3~keV (KK-shell) and 32.3 - 37.3~keV (KL, KM, and KN-shells)~\cite{XENON:2019dti, XENON:2022evz}. Therefore, $^{125}$I ($^{124}$Xe) events in the data are modeled as three (four) Gaussian peaks with relative areas fixed by the capture fractions and widths fixed by the resolution function (Fig.~\ref{fig:energySpectrum}), but the overall normalization as a free fit parameter.

\begin{table}[tbp]
    \footnotesize
    \caption{Expected and fitted background and signal events in the ROI. Note that the fitted uncertainties contains both the statistical and some systematic components. See text for details.}
    \label{tab:solar_nu_background}
    \doublerulesep 0.05pt 
    \tabcolsep 12pt
    \begin{tabular}{l c c} 
    
    \toprule[0.8pt]
    Components & \makecell{Expected \\ events} &\makecell{ Fitted \\events}  \\
     \midrule[0.8pt]
    $^{214}$Pb & 1865 $\pm$ 110 & 1849 $\pm$ 113 \\
    $^{212}$Pb & 276 $\pm$ 71 & 271 $\pm$ 80 \\
    $^{85}$Kr & 489 $\pm$ 254 & 423 $\pm$ 249  \\
    Materials & 683 $\pm$ 27 &  682 $\pm$ 27 \\ 
    $^{136}$Xe & 1009 $\pm$ 46 & 1002 $\pm$ 47\\
    $^{133}$Xe & free & 4767 $\pm$ 135\\
    $^{124}$Xe & free & 317 $\pm$ 63 \\
    $^{125}$I & free & 31 $\pm$ 57 \\
    $^{127}$Xe & free & 59 $\pm$ 23 \\
    \textbf{$pp$+$^{7}$Be neutrino}  & \textbf{-} & \textbf{231 $\pm$ 257}\\
    \bottomrule[0.8pt]
    \end{tabular}
\end{table}

The spectrum fit is based on a binned likelihood procedure using the RooFit package~\cite{Verkerke:2003ir}.
Each background component is simulated using BambooMC~\cite{ Chen:2021asx}, a Monte Carlo simulation package based on the Geant4 framework~\cite{GEANT4:2002zbu}. 
The contributions from $^{214}$Pb, $^{212}$Pb, $^{85}$Kr, detector materials, and $^{136}$Xe are constrained using the uncertainties described earlier and listed in Table~\ref{tab:solar_nu_background}. 
The normalizations of $^{133}$Xe, $^{127}$Xe, $^{125}$I, and $^{124}$Xe are kept free in the fit. 
The sum of solar $pp$ and $^7$Be neutrino signals are combined into a single fit component.

\begin{figure}[tbp]
    \centering
    \includegraphics[width=\columnwidth]{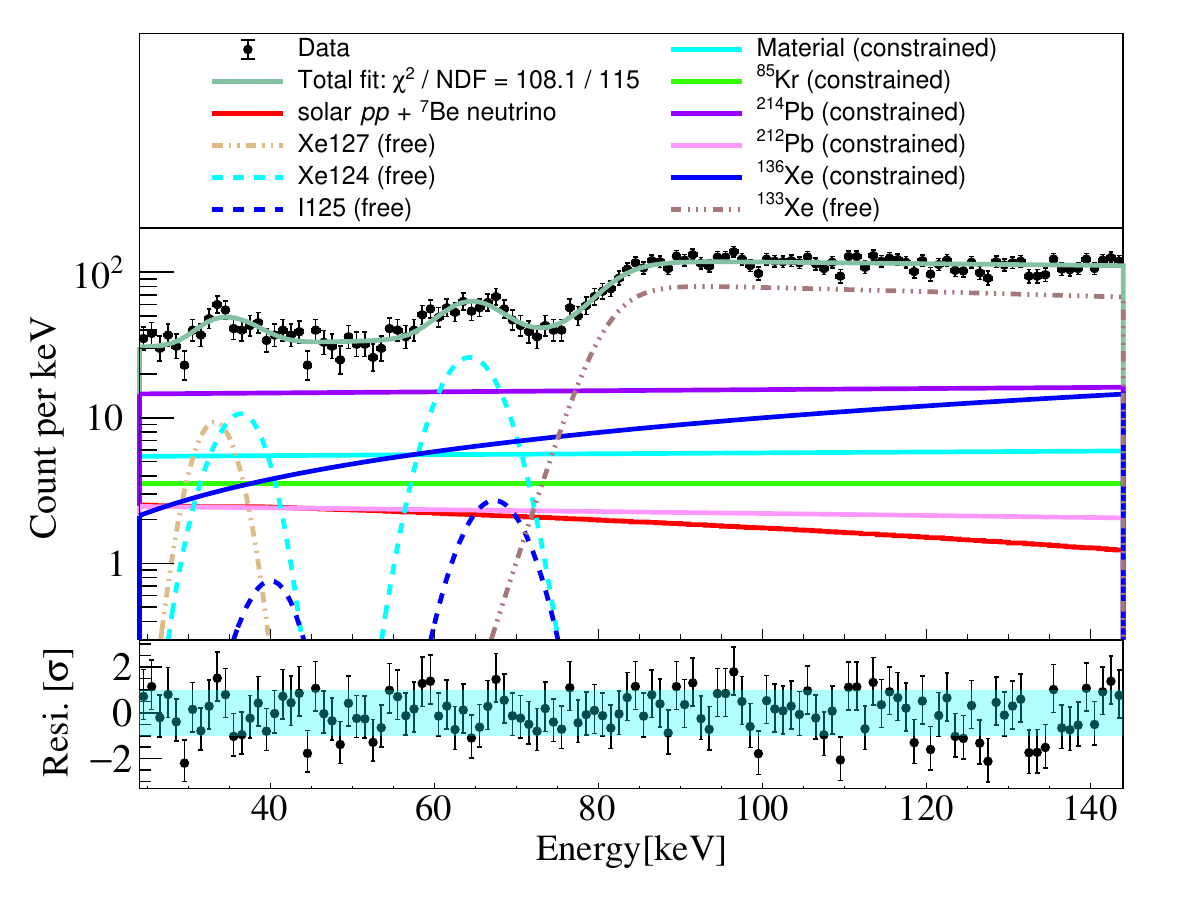}
    \caption{(Color online) Result of the spectrum fit and the corresponding residual plot in the bottom panel. The solar $pp$+$^{7}$Be neutrino is shown as the red line. 
    The constrained and free-floating backgrounds are shown in solid and dashed lines, respectively.
    }
     \label{fig:fig_fitlog}
\end{figure}

The result of the spectrum fit is shown in Fig.~\ref{fig:fig_fitlog}.
The number of solar $pp$ + $^7$Be neutrino signals is 231 $\pm$ 257 events.
Consistent results are obtained with a binned likelihood fit similar to that implemented in Ref.~\cite{PandaX:2022kwg}.
The fitted results of each background are listed in Table~\ref{tab:solar_nu_background}. 
The best-fit background contributions from detector materials, $^{85}$Kr, $^{214}$Pb, and $^{212}$Pb are very close to the input nominal values, indicating that the spectrum fit does not try to drive these background levels up and down, since their shape characters are similar.

The total uncertainty of the $pp$ + $^7$Be neutrino signals is broken down into three terms (Table~\ref{tab:system error}), the statistical uncertainty ($\sigma_{\rm{stat}}$), the systematic uncertainty incorporated in the likelihood fit with nuisance parameters ($\sigma_{\rm{sys1}}$), and additional systematic uncertainty evaluated by hand ($\sigma_{\rm{sys2}}$). We shall discuss them in turn below.

The statistical uncertainty $\sigma_{\rm{stat}}$ is determined to be 113 events, by re-fitting the data with all constrained parameters fixed to their best-fit values without the penalty terms.
The overall size of $\sigma_{\rm{sys1}}$ is then evaluated as $\sqrt{\sigma^2-\sigma_{\rm{stat}}^2}$ to be 231 events, representing the residual component of the fit uncertainty. To quantify the contribution to $\sigma_{\rm{sys1}}$ due to a given component (assuming that it is uncorrelated with the rest), we use a similar approach by refitting the data by only ``tuning on'' the nuisance parameter of this component, and all other nuisance parameters fixed to their best-fit values. The resulting fit uncertainty is then subtracted by $\sigma_{\rm{stat}}$ in quadrature. Individual contributions thus evaluated are also summarized in Table~\ref{tab:system error}. Among all constrained backgrounds, $^{85}$Kr, $^{214}$Pb and $^{212}$Pb dominate the contribution to $\sigma_{\rm{sys1}}$ with an uncertainty of 202, 87, and 69 events, respectively.
The uncertainties of data selection, including the FV, quality cuts, and the SS fraction, affect the signal number proportionally with an estimated uncertainty of 29 events.

Additional systematic uncertainties $\sigma_{\rm{sys2}}$ are evaluated manually. 
The uncertainty from the reconstructed energy scale is determined by comparing the baseline fit result with new fit results, obtained by shifting the data energy spectrum by $\pm1$~keV.
The larger difference led by the shifts is 142, which is used as the uncertainty.
To evaluate the contribution due to the energy resolution uncertainty, we perform the fit by imposing different background spectra with different energy resolution.
A $\pm 1\sigma$ difference in the energy resolution function propagates into an uncertainty of 19 events.
The fit range uncertainty is evaluated to be 29 events by varying the ROI to 25 -- 143~keV and 23 -- 145~keV.
The uncertainty introduced by the $^{214}$Pb spectrum shape estimated to be 84 events, by comparing the theoretical spectrum in the baseline fit to the shape directly measured in the calibration data. 
Similarly, the spectral shape differences between the Geant4~\cite{Chen:2021asx} prediction and a recent theoretical calculation of $^{212}$Pb and $^{85}$Kr~\cite{Haselschwardt:2020iey} are also considered, and the systematic uncertainties are found to be 18 and 5, respectively.
We also fit the data by taking the half-life of $^{136}$Xe $2\nu\beta\beta$ from EXO-200~\cite{EXO-200:2013xfn}, the most precise result so far, as the input value. 
This results in a difference of 16 events in comparison to our baseline fit, which is also taken as a component of systematic uncertainty. In total, $\sigma_{\rm{sys2}}$ is 170 events.

\begin{table}[tbp]
    \footnotesize
    \caption{Summary of contributions to the $pp$ + $^7$Be neutrino signal uncertainties. The contribution of each constrained parameter is extracted by turning on/off the corresponding penalty term in the fitter, assuming no correlation with other constraints. As a result, the quadrature combination of those rows is slightly different from the subtotal $\sigma_{\rm{sys1}}$ obtained directly by the fitter with all correlations properly built in.}
    \label{tab:system error}
    \doublerulesep 0.06pt 
    \tabcolsep 15pt
    \begin{tabular}{l c c} 
    \toprule[0.8pt]
    & Components  & Counts  \\
     \midrule[0.8pt]
    $\sigma_{\rm{stat}}$ & - & 113 \\

     \hline
    & $^{85}$Kr   & 202 \\
      & $^{214}$Pb   & 87 \\
      & $^{212}$Pb & 69 \\ 
      $\sigma_{\rm{sys1}}$ & Material & 21 \\
      & $^{136}$Xe & 19 \\
    & Data selection & 29 \\
    & \textbf{Subtotal} & \textbf{231} \\
    \hline
    & Energy scale & 142 \\
     & Energy resolution & 19  \\
    & Fit range & 29  \\
    $\sigma_{\rm{sys2}}$ & $^{214}$Pb spectrum & 84 \\
    & $^{212}$Pb spectrum & 18 \\
    & $^{85}$Kr spectrum & 5 \\
    & $^{136}$Xe $2\nu\beta\beta$ half-life & 16 \\
    & \textbf{Subtotal} & \textbf{170} \\
    \hline
    & \textbf{Total} & \textbf{287}  \\
    \bottomrule[0.8pt]
    \end{tabular}
\end{table}

\begin{figure}[tbp]
    \centering

    \includegraphics[width=\columnwidth]{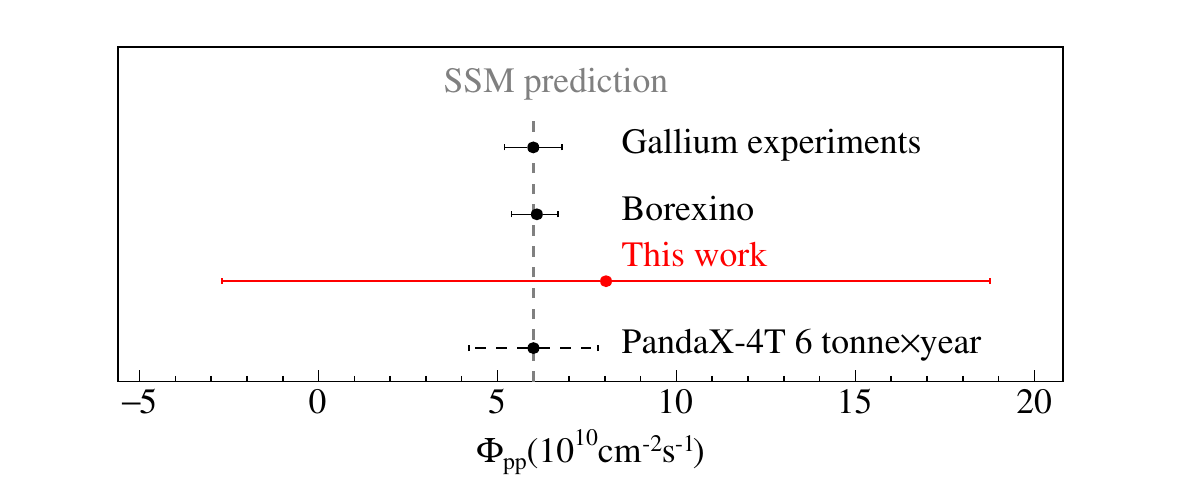}
    \caption{The solar $pp$ neutrino flux measured by PandaX-4T (red), in comparison to those from gallium experiments~\cite{SAGE:2009eeu} and Borexino~\cite{BOREXINO:2018ohr}.
    The vertical dashed line is the average $pp$ neutrino flux from the high-metallicity and low-metallicity SSM~\cite{Vinyoles:2016djt}.
    The PandaX-4T projection with 6 tonne$\times$year exposure is indicated at the bottom (centered at the SSM).
}
    \label{fig:fig5_fit}
\end{figure}

Therefore, the final detected number of $pp$ + $^{7}$Be neutrinos is $231 \pm 113\,(\rm{stat}) \pm 287\,(\rm{syst})$ events from our analysis.
The corresponding solar $pp$ neutrino flux is $(8.0 \pm 3.9\,(\rm{stat}) \pm 10.0\,(\rm{syst})) \times 10^{10}\, \rm{s}^{-1} \rm{cm}^{-2}$, using the expected ratio of the $pp$ to $^{7}$Be fluxes from SSM and their contributions to the ROI.
The result is consistent with the SSM expectation~\cite{Vinyoles:2016djt} and the Borexino measurement~\cite{BOREXINO:2018ohr}, as shown in Fig.~\ref{fig:fig5_fit}.
Based on this, we obtain a $pp$ neutrino flux upper limit of $23.3\times 10^{10}\, $$\rm{s}^{-1} \rm{cm}^{-2}$ at 90\% C.L..

This analysis represents the first result to directly measure the solar $pp$ neutrinos at an electron recoil energy below 150 keV. It demonstrates the potential of multi-tonne-scale LXe detectors for solar neutrino studies in a completely new energy window. 
With an optimized online cryogenic distillation system, we expect to reduce the radon and krypton concentration by a factor of 1.8 or more~\cite{Cui:2020bwf}.
Additional efforts, such as replacing TPC materials and circulation pumps, are being implemented for background reduction.
Additional $^{222}$Rn calibration data and improved detector response with upgraded PMT readout circuit boards are expected to significantly reduce systematic uncertainties. 
Assuming a $^{222}$Rn level of 3.5~$\mu$Bq/kg, a krypton-to-xenon mole concentration of 0.25~ppt, and no short-lived xenon isotopes being activated, and imposing a 5\% constraint on all backgrounds, PandaX-4T can measure the solar $pp$ neutrinos with an uncertainty of 30\% with 6 tonne$\times$year exposure.

This project is supported in part by the grants from National Science Foundation of China (Nos. 12090060, 12090063, 12105052, 12005131, 11905128, 11925502),  the Office of Science and Technology, Shanghai Municipal Government (grant No. 22JC1410100),
and National Science Foundation of Sichuan Province (No. 2024NSFSC1371). 
We thank the support from the Double First Class Plan of Shanghai Jiao Tong University. 
We also thank the sponsorship from the Chinese Academy of Sciences Center for Excellence in Particle Physics (CCEPP), Hongwen Foundation in Hong Kong, Tencent, and New Cornerstone Science Foundation in China. Finally, we thank the CJPL administration and the Yalong River Hydropower Development Company Ltd. for indispensable logistical support and other help.
\bibliography{main.bbl}

\begin{thebibliography}{34}%
\makeatletter
\providecommand \@ifxundefined [1]{%
 \@ifx{#1\undefined}
}%
\providecommand \@ifnum [1]{%
 \ifnum #1\expandafter \@firstoftwo
 \else \expandafter \@secondoftwo
 \fi
}%
\providecommand \@ifx [1]{%
 \ifx #1\expandafter \@firstoftwo
 \else \expandafter \@secondoftwo
 \fi
}%
\providecommand \natexlab [1]{#1}%
\providecommand \enquote  [1]{``#1''}%
\providecommand \bibnamefont  [1]{#1}%
\providecommand \bibfnamefont [1]{#1}%
\providecommand \citenamefont [1]{#1}%
\providecommand \href@noop [0]{\@secondoftwo}%
\providecommand \href [0]{\begingroup \@sanitize@url \@href}%
\providecommand \@href[1]{\@@startlink{#1}\@@href}%
\providecommand \@@href[1]{\endgroup#1\@@endlink}%
\providecommand \@sanitize@url [0]{\catcode `\\12\catcode `\$12\catcode
  `\&12\catcode `\#12\catcode `\^12\catcode `\_12\catcode `\%12\relax}%
\providecommand \@@startlink[1]{}%
\providecommand \@@endlink[0]{}%
\providecommand \url  [0]{\begingroup\@sanitize@url \@url }%
\providecommand \@url [1]{\endgroup\@href {#1}{\urlprefix }}%
\providecommand \urlprefix  [0]{URL }%
\providecommand \Eprint [0]{\href }%
\providecommand \doibase [0]{https://doi.org/}%
\providecommand \selectlanguage [0]{\@gobble}%
\providecommand \bibinfo  [0]{\@secondoftwo}%
\providecommand \bibfield  [0]{\@secondoftwo}%
\providecommand \translation [1]{[#1]}%
\providecommand \BibitemOpen [0]{}%
\providecommand \bibitemStop [0]{}%
\providecommand \bibitemNoStop [0]{.\EOS\space}%
\providecommand \EOS [0]{\spacefactor3000\relax}%
\providecommand \BibitemShut  [1]{\csname bibitem#1\endcsname}%
\let\auto@bib@innerbib\@empty
\bibitem [{\citenamefont {Adelberger}\ \emph {et~al.}(1998)\citenamefont
  {Adelberger} \emph {et~al.}}]{Adelberger:1998qm}%
  \BibitemOpen
  \bibfield  {author} {\bibinfo {author} {\bibfnamefont {E.~G.}\ \bibnamefont
  {Adelberger}} \emph {et~al.},\ }\bibfield  {title} {\bibinfo {title} {{Solar
  fusion cross-sections}},\ }\href {https://doi.org/10.1103/RevModPhys.70.1265}
  {\bibfield  {journal} {\bibinfo  {journal} {Rev. Mod. Phys.}\ }\textbf
  {\bibinfo {volume} {70}},\ \bibinfo {pages} {1265} (\bibinfo {year}
  {1998})},\ \Eprint {https://arxiv.org/abs/astro-ph/9805121}
  {arXiv:astro-ph/9805121} \BibitemShut {NoStop}%
\bibitem [{\citenamefont {Turck-Chieze}\ \emph {et~al.}(2001)\citenamefont
  {Turck-Chieze} \emph {et~al.}}]{Turck-Chieze:2001aug}%
  \BibitemOpen
  \bibfield  {author} {\bibinfo {author} {\bibfnamefont {S.}~\bibnamefont
  {Turck-Chieze}} \emph {et~al.},\ }\bibfield  {title} {\bibinfo {title}
  {{Solar neutrino emission deduced from a seismic model}},\ }\href
  {https://doi.org/10.1086/321726} {\bibfield  {journal} {\bibinfo  {journal}
  {Astrophys. J. Lett.}\ }\textbf {\bibinfo {volume} {555}},\ \bibinfo {pages}
  {L69} (\bibinfo {year} {2001})}\BibitemShut {NoStop}%
\bibitem [{\citenamefont {Adelberger}\ \emph {et~al.}(2011)\citenamefont
  {Adelberger} \emph {et~al.}}]{Adelberger:2010qa}%
  \BibitemOpen
  \bibfield  {author} {\bibinfo {author} {\bibfnamefont {E.~G.}\ \bibnamefont
  {Adelberger}} \emph {et~al.},\ }\bibfield  {title} {\bibinfo {title} {{Solar
  fusion cross sections II: the pp chain and CNO cycles}},\ }\href
  {https://doi.org/10.1103/RevModPhys.83.195} {\bibfield  {journal} {\bibinfo
  {journal} {Rev. Mod. Phys.}\ }\textbf {\bibinfo {volume} {83}},\ \bibinfo
  {pages} {195} (\bibinfo {year} {2011})},\ \Eprint
  {https://arxiv.org/abs/1004.2318} {arXiv:1004.2318 [nucl-ex]} \BibitemShut
  {NoStop}%
\bibitem [{\citenamefont {Vinyoles}\ \emph {et~al.}(2017)\citenamefont
  {Vinyoles}, \citenamefont {Serenelli}, \citenamefont {Villante},
  \citenamefont {Basu}, \citenamefont {Bergstr\"om}, \citenamefont
  {Gonzalez-Garcia}, \citenamefont {Maltoni}, \citenamefont {Pe\~na Garay},\
  and\ \citenamefont {Song}}]{Vinyoles:2016djt}%
  \BibitemOpen
  \bibfield  {author} {\bibinfo {author} {\bibfnamefont {N.}~\bibnamefont
  {Vinyoles}}, \bibinfo {author} {\bibfnamefont {A.~M.}\ \bibnamefont
  {Serenelli}}, \bibinfo {author} {\bibfnamefont {F.~L.}\ \bibnamefont
  {Villante}}, \bibinfo {author} {\bibfnamefont {S.}~\bibnamefont {Basu}},
  \bibinfo {author} {\bibfnamefont {J.}~\bibnamefont {Bergstr\"om}}, \bibinfo
  {author} {\bibfnamefont {M.~C.}\ \bibnamefont {Gonzalez-Garcia}}, \bibinfo
  {author} {\bibfnamefont {M.}~\bibnamefont {Maltoni}}, \bibinfo {author}
  {\bibfnamefont {C.}~\bibnamefont {Pe\~na Garay}},\ and\ \bibinfo {author}
  {\bibfnamefont {N.}~\bibnamefont {Song}},\ }\bibfield  {title} {\bibinfo
  {title} {{A new Generation of Standard Solar Models}},\ }\href
  {https://doi.org/10.3847/1538-4357/835/2/202} {\bibfield  {journal} {\bibinfo
   {journal} {Astrophys. J.}\ }\textbf {\bibinfo {volume} {835}},\ \bibinfo
  {pages} {202} (\bibinfo {year} {2017})},\ \Eprint
  {https://arxiv.org/abs/1611.09867} {arXiv:1611.09867 [astro-ph.SR]}
  \BibitemShut {NoStop}%
\bibitem [{\citenamefont {de~Holanda}\ and\ \citenamefont
  {Smirnov}(2003)}]{deHolanda:2002dko}%
  \BibitemOpen
  \bibfield  {author} {\bibinfo {author} {\bibfnamefont {P.~C.}\ \bibnamefont
  {de~Holanda}}\ and\ \bibinfo {author} {\bibfnamefont {A.~Y.}\ \bibnamefont
  {Smirnov}},\ }\bibfield  {title} {\bibinfo {title} {{LMA MSW solution of the
  solar neutrino problem and first KamLAND results}},\ }\href
  {https://doi.org/10.1088/1475-7516/2003/02/001} {\bibfield  {journal}
  {\bibinfo  {journal} {JCAP}\ }\textbf {\bibinfo {volume} {02}},\ \bibinfo
  {pages} {001}},\ \Eprint {https://arxiv.org/abs/hep-ph/0212270}
  {arXiv:hep-ph/0212270} \BibitemShut {NoStop}%
\bibitem [{\citenamefont {Anselmann}\ \emph {et~al.}(1992)\citenamefont
  {Anselmann} \emph {et~al.}}]{GALLEX:1992gcp}%
  \BibitemOpen
  \bibfield  {author} {\bibinfo {author} {\bibfnamefont {P.}~\bibnamefont
  {Anselmann}} \emph {et~al.} (\bibinfo {collaboration} {GALLEX}),\ }\bibfield
  {title} {\bibinfo {title} {{Solar neutrinos observed by GALLEX at Gran
  Sasso.}},\ }\href {https://doi.org/10.1016/0370-2693(92)91521-A} {\bibfield
  {journal} {\bibinfo  {journal} {Phys. Lett. B}\ }\textbf {\bibinfo {volume}
  {285}},\ \bibinfo {pages} {376} (\bibinfo {year} {1992})}\BibitemShut
  {NoStop}%
\bibitem [{\citenamefont {Abdurashitov}\ \emph {et~al.}(1994)\citenamefont
  {Abdurashitov} \emph {et~al.}}]{SAGE:1994ctc}%
  \BibitemOpen
  \bibfield  {author} {\bibinfo {author} {\bibfnamefont {J.~N.}\ \bibnamefont
  {Abdurashitov}} \emph {et~al.} (\bibinfo {collaboration} {SAGE}),\ }\bibfield
   {title} {\bibinfo {title} {{Results from SAGE}},\ }\href
  {https://doi.org/10.1016/0370-2693(94)90454-5} {\bibfield  {journal}
  {\bibinfo  {journal} {Phys. Lett. B}\ }\textbf {\bibinfo {volume} {328}},\
  \bibinfo {pages} {234} (\bibinfo {year} {1994})}\BibitemShut {NoStop}%
\bibitem [{\citenamefont {Bellini}\ \emph {et~al.}(2014)\citenamefont {Bellini}
  \emph {et~al.}}]{BOREXINO:2014pcl}%
  \BibitemOpen
  \bibfield  {author} {\bibinfo {author} {\bibfnamefont {G.}~\bibnamefont
  {Bellini}} \emph {et~al.} (\bibinfo {collaboration} {BOREXINO}),\ }\bibfield
  {title} {\bibinfo {title} {{Neutrinos from the primary
  proton\textendash{}proton fusion process in the Sun}},\ }\href
  {https://doi.org/10.1038/nature13702} {\bibfield  {journal} {\bibinfo
  {journal} {Nature}\ }\textbf {\bibinfo {volume} {512}},\ \bibinfo {pages}
  {383} (\bibinfo {year} {2014})}\BibitemShut {NoStop}%
\bibitem [{\citenamefont {Agostini}\ \emph {et~al.}(2018)\citenamefont
  {Agostini} \emph {et~al.}}]{BOREXINO:2018ohr}%
  \BibitemOpen
  \bibfield  {author} {\bibinfo {author} {\bibfnamefont {M.}~\bibnamefont
  {Agostini}} \emph {et~al.} (\bibinfo {collaboration} {BOREXINO}),\ }\bibfield
   {title} {\bibinfo {title} {{Comprehensive measurement of $pp$-chain solar
  neutrinos}},\ }\href {https://doi.org/10.1038/s41586-018-0624-y} {\bibfield
  {journal} {\bibinfo  {journal} {Nature}\ }\textbf {\bibinfo {volume} {562}},\
  \bibinfo {pages} {505} (\bibinfo {year} {2018})}\BibitemShut {NoStop}%
\bibitem [{\citenamefont {Meng}\ \emph {et~al.}(2021)\citenamefont {Meng} \emph
  {et~al.}}]{PandaX-4T:2021bab}%
  \BibitemOpen
  \bibfield  {author} {\bibinfo {author} {\bibfnamefont {Y.}~\bibnamefont
  {Meng}} \emph {et~al.} (\bibinfo {collaboration} {PandaX-4T}),\ }\bibfield
  {title} {\bibinfo {title} {{Dark Matter Search Results from the PandaX-4T
  Commissioning Run}},\ }\href {https://doi.org/10.1103/PhysRevLett.127.261802}
  {\bibfield  {journal} {\bibinfo  {journal} {Phys. Rev. Lett.}\ }\textbf
  {\bibinfo {volume} {127}},\ \bibinfo {pages} {261802} (\bibinfo {year}
  {2021})},\ \Eprint {https://arxiv.org/abs/2107.13438} {arXiv:2107.13438
  [hep-ex]} \BibitemShut {NoStop}%
\bibitem [{\citenamefont {Akerib}\ \emph {et~al.}(2020)\citenamefont {Akerib}
  \emph {et~al.}}]{LZ:2018qzl}%
  \BibitemOpen
  \bibfield  {author} {\bibinfo {author} {\bibfnamefont {D.~S.}\ \bibnamefont
  {Akerib}} \emph {et~al.} (\bibinfo {collaboration} {LZ}),\ }\bibfield
  {title} {\bibinfo {title} {{Projected WIMP sensitivity of the LUX-ZEPLIN dark
  matter experiment}},\ }\href {https://doi.org/10.1103/PhysRevD.101.052002}
  {\bibfield  {journal} {\bibinfo  {journal} {Phys. Rev. D}\ }\textbf {\bibinfo
  {volume} {101}},\ \bibinfo {pages} {052002} (\bibinfo {year} {2020})},\
  \Eprint {https://arxiv.org/abs/1802.06039} {arXiv:1802.06039 [astro-ph.IM]}
  \BibitemShut {NoStop}%
\bibitem [{\citenamefont {Aprile}\ \emph
  {et~al.}(2020{\natexlab{a}})\citenamefont {Aprile} \emph
  {et~al.}}]{XENON:2020kmp}%
  \BibitemOpen
  \bibfield  {author} {\bibinfo {author} {\bibfnamefont {E.}~\bibnamefont
  {Aprile}} \emph {et~al.} (\bibinfo {collaboration} {XENON}),\ }\bibfield
  {title} {\bibinfo {title} {{Projected WIMP sensitivity of the XENONnT dark
  matter experiment}},\ }\href {https://doi.org/10.1088/1475-7516/2020/11/031}
  {\bibfield  {journal} {\bibinfo  {journal} {JCAP}\ }\textbf {\bibinfo
  {volume} {11}},\ \bibinfo {pages} {031}},\ \Eprint
  {https://arxiv.org/abs/2007.08796} {arXiv:2007.08796 [physics.ins-det]}
  \BibitemShut {NoStop}%
\bibitem [{\citenamefont {Ma}\ \emph {et~al.}(2023)\citenamefont {Ma} \emph
  {et~al.}}]{PandaX:2022aac}%
  \BibitemOpen
  \bibfield  {author} {\bibinfo {author} {\bibfnamefont {W.}~\bibnamefont {Ma}}
  \emph {et~al.} (\bibinfo {collaboration} {PandaX}),\ }\bibfield  {title}
  {\bibinfo {title} {{Search for Solar B8 Neutrinos in the PandaX-4T Experiment
  Using Neutrino-Nucleus Coherent Scattering}},\ }\href
  {https://doi.org/10.1103/PhysRevLett.130.021802} {\bibfield  {journal}
  {\bibinfo  {journal} {Phys. Rev. Lett.}\ }\textbf {\bibinfo {volume} {130}},\
  \bibinfo {pages} {021802} (\bibinfo {year} {2023})},\ \Eprint
  {https://arxiv.org/abs/2207.04883} {arXiv:2207.04883 [hep-ex]} \BibitemShut
  {NoStop}%
\bibitem [{\citenamefont {Aprile}\ \emph {et~al.}(2021)\citenamefont {Aprile}
  \emph {et~al.}}]{XENON:2020gfr}%
  \BibitemOpen
  \bibfield  {author} {\bibinfo {author} {\bibfnamefont {E.}~\bibnamefont
  {Aprile}} \emph {et~al.} (\bibinfo {collaboration} {XENON}),\ }\bibfield
  {title} {\bibinfo {title} {{Search for Coherent Elastic Scattering of Solar
  $^8$B Neutrinos in the XENON1T Dark Matter Experiment}},\ }\href
  {https://doi.org/10.1103/PhysRevLett.126.091301} {\bibfield  {journal}
  {\bibinfo  {journal} {Phys. Rev. Lett.}\ }\textbf {\bibinfo {volume} {126}},\
  \bibinfo {pages} {091301} (\bibinfo {year} {2021})},\ \Eprint
  {https://arxiv.org/abs/2012.02846} {arXiv:2012.02846 [hep-ex]} \BibitemShut
  {NoStop}%
\bibitem [{\citenamefont {Zhou}\ \emph {et~al.}(2021)\citenamefont {Zhou} \emph
  {et~al.}}]{PandaX-II:2020udv}%
  \BibitemOpen
  \bibfield  {author} {\bibinfo {author} {\bibfnamefont {X.}~\bibnamefont
  {Zhou}} \emph {et~al.} (\bibinfo {collaboration} {PandaX-II}),\ }\bibfield
  {title} {\bibinfo {title} {{A Search for Solar Axions and Anomalous Neutrino
  Magnetic Moment with the Complete PandaX-II Data}},\ }\href
  {https://doi.org/10.1088/0256-307X/38/10/109902} {\bibfield  {journal}
  {\bibinfo  {journal} {Chin. Phys. Lett.}\ }\textbf {\bibinfo {volume} {38}},\
  \bibinfo {pages} {011301} (\bibinfo {year} {2021})},\ \bibinfo {note}
  {[Erratum: Chin.Phys.Lett. 38, 109902 (2021)]},\ \Eprint
  {https://arxiv.org/abs/2008.06485} {arXiv:2008.06485 [hep-ex]} \BibitemShut
  {NoStop}%
\bibitem [{\citenamefont {Aprile}\ \emph
  {et~al.}(2020{\natexlab{b}})\citenamefont {Aprile} \emph
  {et~al.}}]{XENON:2020rca}%
  \BibitemOpen
  \bibfield  {author} {\bibinfo {author} {\bibfnamefont {E.}~\bibnamefont
  {Aprile}} \emph {et~al.} (\bibinfo {collaboration} {XENON}),\ }\bibfield
  {title} {\bibinfo {title} {{Excess electronic recoil events in XENON1T}},\
  }\href {https://doi.org/10.1103/PhysRevD.102.072004} {\bibfield  {journal}
  {\bibinfo  {journal} {Phys. Rev. D}\ }\textbf {\bibinfo {volume} {102}},\
  \bibinfo {pages} {072004} (\bibinfo {year} {2020}{\natexlab{b}})},\ \Eprint
  {https://arxiv.org/abs/2006.09721} {arXiv:2006.09721 [hep-ex]} \BibitemShut
  {NoStop}%
\bibitem [{\citenamefont {Billard}\ \emph {et~al.}(2014)\citenamefont
  {Billard}, \citenamefont {Strigari},\ and\ \citenamefont
  {Figueroa-Feliciano}}]{Billard:2013qya}%
  \BibitemOpen
  \bibfield  {author} {\bibinfo {author} {\bibfnamefont {J.}~\bibnamefont
  {Billard}}, \bibinfo {author} {\bibfnamefont {L.}~\bibnamefont {Strigari}},\
  and\ \bibinfo {author} {\bibfnamefont {E.}~\bibnamefont
  {Figueroa-Feliciano}},\ }\bibfield  {title} {\bibinfo {title} {{Implication
  of neutrino backgrounds on the reach of next generation dark matter direct
  detection experiments}},\ }\href {https://doi.org/10.1103/PhysRevD.89.023524}
  {\bibfield  {journal} {\bibinfo  {journal} {Phys. Rev. D}\ }\textbf {\bibinfo
  {volume} {89}},\ \bibinfo {pages} {023524} (\bibinfo {year} {2014})},\
  \Eprint {https://arxiv.org/abs/1307.5458} {arXiv:1307.5458 [hep-ph]}
  \BibitemShut {NoStop}%
\bibitem [{\citenamefont {Chen}\ \emph {et~al.}(2017)\citenamefont {Chen},
  \citenamefont {Chi}, \citenamefont {Liu},\ and\ \citenamefont
  {Wu}}]{Chen:2016eab}%
  \BibitemOpen
  \bibfield  {author} {\bibinfo {author} {\bibfnamefont {J.-W.}\ \bibnamefont
  {Chen}}, \bibinfo {author} {\bibfnamefont {H.-C.}\ \bibnamefont {Chi}},
  \bibinfo {author} {\bibfnamefont {C.~P.}\ \bibnamefont {Liu}},\ and\ \bibinfo
  {author} {\bibfnamefont {C.-P.}\ \bibnamefont {Wu}},\ }\bibfield  {title}
  {\bibinfo {title} {{Low-energy electronic recoil in xenon detectors by solar
  neutrinos}},\ }\href {https://doi.org/10.1016/j.physletb.2017.10.029}
  {\bibfield  {journal} {\bibinfo  {journal} {Phys. Lett. B}\ }\textbf
  {\bibinfo {volume} {774}},\ \bibinfo {pages} {656} (\bibinfo {year}
  {2017})},\ \Eprint {https://arxiv.org/abs/1610.04177} {arXiv:1610.04177
  [hep-ex]} \BibitemShut {NoStop}%
\bibitem [{\citenamefont {Ma}\ \emph {et~al.}(2020)\citenamefont {Ma} \emph
  {et~al.}}]{CDEX:2017kys}%
  \BibitemOpen
  \bibfield  {author} {\bibinfo {author} {\bibfnamefont {H.}~\bibnamefont {Ma}}
  \emph {et~al.} (\bibinfo {collaboration} {CDEX}),\ }\bibfield  {title}
  {\bibinfo {title} {{CDEX Dark Matter Experiment: Status and Prospects}},\
  }\href {https://doi.org/10.1088/1742-6596/1342/1/012067} {\bibfield
  {journal} {\bibinfo  {journal} {J. Phys. Conf. Ser.}\ }\textbf {\bibinfo
  {volume} {1342}},\ \bibinfo {pages} {012067} (\bibinfo {year} {2020})},\
  \Eprint {https://arxiv.org/abs/1712.06046} {arXiv:1712.06046 [hep-ex]}
  \BibitemShut {NoStop}%
\bibitem [{\citenamefont {Li}\ \emph {et~al.}(2015)\citenamefont {Li},
  \citenamefont {Ji}, \citenamefont {Haxton},\ and\ \citenamefont
  {Wang}}]{Li:2014rca}%
  \BibitemOpen
  \bibfield  {author} {\bibinfo {author} {\bibfnamefont {J.}~\bibnamefont
  {Li}}, \bibinfo {author} {\bibfnamefont {X.}~\bibnamefont {Ji}}, \bibinfo
  {author} {\bibfnamefont {W.}~\bibnamefont {Haxton}},\ and\ \bibinfo {author}
  {\bibfnamefont {J.~S.~Y.}\ \bibnamefont {Wang}},\ }\bibfield  {title}
  {\bibinfo {title} {{The second-phase development of the China JinPing
  underground Laboratory}},\ }\href
  {https://doi.org/10.1016/j.phpro.2014.12.055} {\bibfield  {journal} {\bibinfo
   {journal} {Phys. Procedia}\ }\textbf {\bibinfo {volume} {61}},\ \bibinfo
  {pages} {576} (\bibinfo {year} {2015})},\ \Eprint
  {https://arxiv.org/abs/1404.2651} {arXiv:1404.2651 [physics.ins-det]}
  \BibitemShut {NoStop}%
\bibitem [{\citenamefont {Yan}\ \emph {et~al.}(2024)\citenamefont {Yan} \emph
  {et~al.}}]{PandaX:2023ggs}%
  \BibitemOpen
  \bibfield  {author} {\bibinfo {author} {\bibfnamefont {X.}~\bibnamefont
  {Yan}} \emph {et~al.} (\bibinfo {collaboration} {PandaX}),\ }\bibfield
  {title} {\bibinfo {title} {{Searching for Two-Neutrino and Neutrinoless
  Double Beta Decay of Xe134 with the PandaX-4T Experiment}},\ }\href
  {https://doi.org/10.1103/PhysRevLett.132.152502} {\bibfield  {journal}
  {\bibinfo  {journal} {Phys. Rev. Lett.}\ }\textbf {\bibinfo {volume} {132}},\
  \bibinfo {pages} {152502} (\bibinfo {year} {2024})},\ \Eprint
  {https://arxiv.org/abs/2312.15632} {arXiv:2312.15632 [nucl-ex]} \BibitemShut
  {NoStop}%
\bibitem [{\citenamefont {Si}\ \emph {et~al.}(2022)\citenamefont {Si} \emph
  {et~al.}}]{PandaX:2022kwg}%
  \BibitemOpen
  \bibfield  {author} {\bibinfo {author} {\bibfnamefont {L.}~\bibnamefont {Si}}
  \emph {et~al.} (\bibinfo {collaboration} {PandaX}),\ }\bibfield  {title}
  {\bibinfo {title} {{Determination of Double Beta Decay Half-Life of 136Xe
  with the PandaX-4T Natural Xenon Detector}},\ }\href
  {https://doi.org/10.34133/2022/9798721} {\bibfield  {journal} {\bibinfo
  {journal} {Research}\ }\textbf {\bibinfo {volume} {2022}},\ \bibinfo {pages}
  {9798721} (\bibinfo {year} {2022})},\ \Eprint
  {https://arxiv.org/abs/2205.12809} {arXiv:2205.12809 [nucl-ex]} \BibitemShut
  {NoStop}%
\bibitem [{\citenamefont {Zhang}\ \emph {et~al.}(2022)\citenamefont {Zhang}
  \emph {et~al.}}]{Zhang:2021shp}%
  \BibitemOpen
  \bibfield  {author} {\bibinfo {author} {\bibfnamefont {D.}~\bibnamefont
  {Zhang}} \emph {et~al.},\ }\bibfield  {title} {\bibinfo {title} {{Rb83/Kr83m
  production and cross-section measurement with 3.4 MeV and 20 MeV proton
  beams}},\ }\href {https://doi.org/10.1103/PhysRevC.105.014604} {\bibfield
  {journal} {\bibinfo  {journal} {Phys. Rev. C}\ }\textbf {\bibinfo {volume}
  {105}},\ \bibinfo {pages} {014604} (\bibinfo {year} {2022})},\ \Eprint
  {https://arxiv.org/abs/2102.02490} {arXiv:2102.02490 [nucl-ex]} \BibitemShut
  {NoStop}%
\bibitem [{\citenamefont {Szydagis}\ \emph {et~al.}(2011)\citenamefont
  {Szydagis}, \citenamefont {Barry}, \citenamefont {Kazkaz}, \citenamefont
  {Mock}, \citenamefont {Stolp}, \citenamefont {Sweany}, \citenamefont
  {Tripathi}, \citenamefont {Uvarov}, \citenamefont {Walsh},\ and\
  \citenamefont {Woods}}]{Szydagis:2011tk}%
  \BibitemOpen
  \bibfield  {author} {\bibinfo {author} {\bibfnamefont {M.}~\bibnamefont
  {Szydagis}}, \bibinfo {author} {\bibfnamefont {N.}~\bibnamefont {Barry}},
  \bibinfo {author} {\bibfnamefont {K.}~\bibnamefont {Kazkaz}}, \bibinfo
  {author} {\bibfnamefont {J.}~\bibnamefont {Mock}}, \bibinfo {author}
  {\bibfnamefont {D.}~\bibnamefont {Stolp}}, \bibinfo {author} {\bibfnamefont
  {M.}~\bibnamefont {Sweany}}, \bibinfo {author} {\bibfnamefont
  {M.}~\bibnamefont {Tripathi}}, \bibinfo {author} {\bibfnamefont
  {S.}~\bibnamefont {Uvarov}}, \bibinfo {author} {\bibfnamefont
  {N.}~\bibnamefont {Walsh}},\ and\ \bibinfo {author} {\bibfnamefont
  {M.}~\bibnamefont {Woods}},\ }\bibfield  {title} {\bibinfo {title} {{NEST: A
  Comprehensive Model for Scintillation Yield in Liquid Xenon}},\ }\href
  {https://doi.org/10.1088/1748-0221/6/10/P10002} {\bibfield  {journal}
  {\bibinfo  {journal} {JINST}\ }\textbf {\bibinfo {volume} {6}},\ \bibinfo
  {pages} {P10002}},\ \Eprint {https://arxiv.org/abs/1106.1613}
  {arXiv:1106.1613 [physics.ins-det]} \BibitemShut {NoStop}%
\bibitem [{\citenamefont {Aprile}\ \emph
  {et~al.}(2020{\natexlab{c}})\citenamefont {Aprile} \emph
  {et~al.}}]{XENON:2020iwh}%
  \BibitemOpen
  \bibfield  {author} {\bibinfo {author} {\bibfnamefont {E.}~\bibnamefont
  {Aprile}} \emph {et~al.} (\bibinfo {collaboration} {XENON}),\ }\bibfield
  {title} {\bibinfo {title} {{Energy resolution and linearity of XENON1T in the
  MeV energy range}},\ }\href {https://doi.org/10.1140/epjc/s10052-020-8284-0}
  {\bibfield  {journal} {\bibinfo  {journal} {Eur. Phys. J. C}\ }\textbf
  {\bibinfo {volume} {80}},\ \bibinfo {pages} {785} (\bibinfo {year}
  {2020}{\natexlab{c}})},\ \Eprint {https://arxiv.org/abs/2003.03825}
  {arXiv:2003.03825 [physics.ins-det]} \BibitemShut {NoStop}%
\bibitem [{\citenamefont {Haselschwardt}\ \emph {et~al.}(2020)\citenamefont
  {Haselschwardt}, \citenamefont {Kostensalo}, \citenamefont {Mougeot},\ and\
  \citenamefont {Suhonen}}]{Haselschwardt:2020iey}%
  \BibitemOpen
  \bibfield  {author} {\bibinfo {author} {\bibfnamefont {S.~J.}\ \bibnamefont
  {Haselschwardt}}, \bibinfo {author} {\bibfnamefont {J.}~\bibnamefont
  {Kostensalo}}, \bibinfo {author} {\bibfnamefont {X.}~\bibnamefont
  {Mougeot}},\ and\ \bibinfo {author} {\bibfnamefont {J.}~\bibnamefont
  {Suhonen}},\ }\bibfield  {title} {\bibinfo {title} {{Improved calculations of
  beta decay backgrounds to new physics in liquid xenon detectors}},\ }\href
  {https://doi.org/10.1103/PhysRevC.102.065501} {\bibfield  {journal} {\bibinfo
   {journal} {Phys. Rev. C}\ }\textbf {\bibinfo {volume} {102}},\ \bibinfo
  {pages} {065501} (\bibinfo {year} {2020})},\ \Eprint
  {https://arxiv.org/abs/2007.13686} {arXiv:2007.13686 [hep-ex]} \BibitemShut
  {NoStop}%
\bibitem [{\citenamefont {Aprile}\ \emph {et~al.}(2022)\citenamefont {Aprile}
  \emph {et~al.}}]{XENON:2022evz}%
  \BibitemOpen
  \bibfield  {author} {\bibinfo {author} {\bibfnamefont {E.}~\bibnamefont
  {Aprile}} \emph {et~al.} (\bibinfo {collaboration} {XENON}),\ }\bibfield
  {title} {\bibinfo {title} {{Double-Weak Decays of $^{124}$Xe and $^{136}$Xe
  in the XENON1T and XENONnT Experiments}},\ }\href
  {https://doi.org/10.1103/PhysRevC.106.024328} {\bibfield  {journal} {\bibinfo
   {journal} {Phys. Rev. C}\ }\textbf {\bibinfo {volume} {106}},\ \bibinfo
  {pages} {024328} (\bibinfo {year} {2022})},\ \Eprint
  {https://arxiv.org/abs/2205.04158} {arXiv:2205.04158 [hep-ex]} \BibitemShut
  {NoStop}%
\bibitem [{\citenamefont {Aprile}\ \emph {et~al.}(2019)\citenamefont {Aprile}
  \emph {et~al.}}]{XENON:2019dti}%
  \BibitemOpen
  \bibfield  {author} {\bibinfo {author} {\bibfnamefont {E.}~\bibnamefont
  {Aprile}} \emph {et~al.} (\bibinfo {collaboration} {XENON}),\ }\bibfield
  {title} {\bibinfo {title} {{Observation of two-neutrino double electron
  capture in $^{124}$Xe with XENON1T}},\ }\href
  {https://doi.org/10.1038/s41586-019-1124-4} {\bibfield  {journal} {\bibinfo
  {journal} {Nature}\ }\textbf {\bibinfo {volume} {568}},\ \bibinfo {pages}
  {532} (\bibinfo {year} {2019})},\ \Eprint {https://arxiv.org/abs/1904.11002}
  {arXiv:1904.11002 [nucl-ex]} \BibitemShut {NoStop}%
\bibitem [{\citenamefont {Verkerke}\ and\ \citenamefont
  {Kirkby}(2003)}]{Verkerke:2003ir}%
  \BibitemOpen
  \bibfield  {author} {\bibinfo {author} {\bibfnamefont {W.}~\bibnamefont
  {Verkerke}}\ and\ \bibinfo {author} {\bibfnamefont {D.~P.}\ \bibnamefont
  {Kirkby}},\ }\bibfield  {title} {\bibinfo {title} {{The RooFit toolkit for
  data modeling}},\ }\href@noop {} {\bibfield  {journal} {\bibinfo  {journal}
  {eConf}\ }\textbf {\bibinfo {volume} {C0303241}},\ \bibinfo {pages} {MOLT007}
  (\bibinfo {year} {2003})},\ \Eprint {https://arxiv.org/abs/physics/0306116}
  {arXiv:physics/0306116} \BibitemShut {NoStop}%
\bibitem [{\citenamefont {Chen}\ \emph {et~al.}(2021)\citenamefont {Chen} \emph
  {et~al.}}]{Chen:2021asx}%
  \BibitemOpen
  \bibfield  {author} {\bibinfo {author} {\bibfnamefont {X.}~\bibnamefont
  {Chen}} \emph {et~al.},\ }\bibfield  {title} {\bibinfo {title} {{BambooMC
  \textemdash{} A Geant4-based simulation program for the PandaX
  experiments}},\ }\href {https://doi.org/10.1088/1748-0221/16/09/T09004}
  {\bibfield  {journal} {\bibinfo  {journal} {JINST}\ }\textbf {\bibinfo
  {volume} {16}}\bibfield  {number} {\bibinfo  {number} { (09)},\ \bibinfo
  {pages} {T09004}},\ }\Eprint {https://arxiv.org/abs/2107.05935}
  {arXiv:2107.05935 [physics.ins-det]} \BibitemShut {NoStop}%
\bibitem [{\citenamefont {Agostinelli}\ \emph {et~al.}(2003)\citenamefont
  {Agostinelli} \emph {et~al.}}]{GEANT4:2002zbu}%
  \BibitemOpen
  \bibfield  {author} {\bibinfo {author} {\bibfnamefont {S.}~\bibnamefont
  {Agostinelli}} \emph {et~al.} (\bibinfo {collaboration} {GEANT4}),\
  }\bibfield  {title} {\bibinfo {title} {{GEANT4--a simulation toolkit}},\
  }\href {https://doi.org/10.1016/S0168-9002(03)01368-8} {\bibfield  {journal}
  {\bibinfo  {journal} {Nucl. Instrum. Meth. A}\ }\textbf {\bibinfo {volume}
  {506}},\ \bibinfo {pages} {250} (\bibinfo {year} {2003})}\BibitemShut
  {NoStop}%
\bibitem [{\citenamefont {Albert}\ \emph {et~al.}(2014)\citenamefont {Albert}
  \emph {et~al.}}]{EXO-200:2013xfn}%
  \BibitemOpen
  \bibfield  {author} {\bibinfo {author} {\bibfnamefont {J.~B.}\ \bibnamefont
  {Albert}} \emph {et~al.} (\bibinfo {collaboration} {EXO-200}),\ }\bibfield
  {title} {\bibinfo {title} {{Improved measurement of the $2\nu\beta\beta$
  half-life of $^{136}$Xe with the EXO-200 detector}},\ }\href
  {https://doi.org/10.1103/PhysRevC.89.015502} {\bibfield  {journal} {\bibinfo
  {journal} {Phys. Rev. C}\ }\textbf {\bibinfo {volume} {89}},\ \bibinfo
  {pages} {015502} (\bibinfo {year} {2014})},\ \Eprint
  {https://arxiv.org/abs/1306.6106} {arXiv:1306.6106 [nucl-ex]} \BibitemShut
  {NoStop}%
\bibitem [{\citenamefont {Cui}\ \emph {et~al.}(2021)\citenamefont {Cui} \emph
  {et~al.}}]{Cui:2020bwf}%
  \BibitemOpen
  \bibfield  {author} {\bibinfo {author} {\bibfnamefont {X.}~\bibnamefont
  {Cui}} \emph {et~al.},\ }\bibfield  {title} {\bibinfo {title} {{Design and
  commissioning of the PandaX-4T cryogenic distillation system for krypton and
  radon removal}},\ }\href {https://doi.org/10.1088/1748-0221/16/07/P07046}
  {\bibfield  {journal} {\bibinfo  {journal} {JINST}\ }\textbf {\bibinfo
  {volume} {16}}\bibfield  {number} {\bibinfo  {number} { (07)},\ \bibinfo
  {pages} {P07046}},\ }\Eprint {https://arxiv.org/abs/2012.02436}
  {arXiv:2012.02436 [physics.ins-det]} \BibitemShut {NoStop}%
\bibitem [{\citenamefont {Abdurashitov}\ \emph {et~al.}(2009)\citenamefont
  {Abdurashitov} \emph {et~al.}}]{SAGE:2009eeu}%
  \BibitemOpen
  \bibfield  {author} {\bibinfo {author} {\bibfnamefont {J.~N.}\ \bibnamefont
  {Abdurashitov}} \emph {et~al.} (\bibinfo {collaboration} {SAGE}),\ }\bibfield
   {title} {\bibinfo {title} {{Measurement of the solar neutrino capture rate
  with gallium metal. III: Results for the 2002--2007 data-taking period}},\
  }\href {https://doi.org/10.1103/PhysRevC.80.015807} {\bibfield  {journal}
  {\bibinfo  {journal} {Phys. Rev. C}\ }\textbf {\bibinfo {volume} {80}},\
  \bibinfo {pages} {015807} (\bibinfo {year} {2009})},\ \Eprint
  {https://arxiv.org/abs/0901.2200} {arXiv:0901.2200 [nucl-ex]} \BibitemShut
  {NoStop}%
\end{thebibliography}%
\end{document}